\documentstyle[11pt,newpasp,twoside,epsf]{article} 
\def\edcomment#1{\iffalse\marginpar{\raggedright\sl#1\/}\else\relax\fi} 
\reversemarginpar 

\begin{document} 
\title{Hot Stellar Populations in Globular Clusters:\\ a Photometrist's View}

\author{Giampaolo Piotto} 
\affil{Dipartimento di Astronomia, Universit\`a di Padova, Vicolo dell'Osservatorio, 2, 
I-35122 Padova, Italy} 

\begin{abstract} 
We briefly review the recent results on hot horizontal-branch stars in
globular clusters. Since the first Ivanfest, in 1992, there have been
a number of new observational lines of evidence which have allowed 
significant progress in our understanding of blue-tail stars, though
new, even more intriguing questions arise. Despite this
progress, we still do not know the answer to the main question: {\it
why} are there blue-HB-tail stars? 

The new photometric data bases collected in the last few years, and the
forthcoming multi-fiber observational campaigns on 10m-class telescopes,
might be the key to solve this puzzle. We will show an example of how 
these data bases can disclose important properties of blue HB stars.
\end{abstract}

\section{Introduction} 

Horizontal branches (HB) are probably the most challenging structures
in the color magnitude diagrams (CMD) of old stellar
populations. Despite the fact that all the HB stars have almost the same
absolute luminosity and are at the same evolutionary stage, a simple look at
a sample of observed globular-cluster (GC) CMDs (e.g., the HST snapshot
survey in Piotto et al.\ 2002) shows a large variety of morphologies,
suggesting immediately a complex scenario. As originally realized
about 40 years ago by Sandage and Wallerstein (1960), Faulkner (1996),
and van den Bergh (1967), the location of a star on the HB depends on
a large number of parameters. Indeed, we now know that the temperature
of a star on the zero age HB (ZAHB) depends on {\it all} the
parameters involved in the stellar models. On the one hand, this fact
makes the interpretation of the HB morphologies rather difficult. On
the other hand, as the HBs behave as a sort of amplifiers of the stellar
population properties, they are an ideal laboratory in which to study the
initial conditions and the evolution of the single stars as well as of
the parent cluster. Understanding the HB morphology has a
large astrophysical impact. The HBs allow us to understand the evolution of
Population II stars, and of specific systems (clusters or nearby
galaxies). Moreover, as HB stars are bright (particularly at short
wavelengths), they are key ingredients in any stellar-population
synthesis models; and therefore the knowledge of their properties has
a significant impact on understanding the evolution of distant, unresolved
galaxies. If we note  that HB stars (in particular the RR Lyrae) play
a fundamental role in the calibration of the cosmic distance
scale, we see that the study of the HBs is relevant also for
cosmology.

Despite the frustration one might sometimes feel when confronted with the zoo
of HB properties, the large effort devoted to their study 
is undoubtedly worth its while.

\section{The Open Problems}

A number of observational properties of HBs are not well understood,
despite the fact that they have been known for more than 30 years. Among them
we mention: \\
(1) The Oosterhoff-Sandage period shift, concerning the pulsation properties
of the RR Lyrae variables (Oosterhoff 1939, Arp 1955, Sandage 1981);\\
(2) The dependence of the HB morphology on metallicity ;\\
(3) Horizontal-branch bimodality (Harris 1974);\\
(4) The non-monotonic correlation between the HB extension and the metallicity,
i.e., the evidence that there must be a ``second parameter'' 
(Sandage and Wildey 1967), or a combination of parameters (Buonanno et
al.\ 1997) that determine the observed HB morphology;\\
(5) The ``tilted HB'' (Raimondo et al.\ 2002), i.e., a dependence of the HB luminosity
on the temperature (cf.\ also Raimondo et al.\ in this book);\\
(6) The presence of blue tails (Fusi Pecci et al.\ 1993).

In the last ten years, since the first Ivanfest, most of the
investigation of globular cluster HBs has been devoted to the blue
tails (BTs), and we will concentrate on them in the remainder of this
paper.  The BT is an extension of the HB that runs almost vertically
in the classical $V$ vs.\ $B-V$ (or $V$ vs.\ $V-I$) CMD, because of the
insensitivity of the $B-V$ (or $V-I$) color to the temperature, for
$T_e\ge 10,000$~K. The BT has a different morphology in different
bandpasses. In the literature, there is some confusion concerning the
name of the different branches of the hot HB; in the following, we will
refer as BT to the part of the HB beyond $T_e=10,000$~K, which
corresponds observationally to the region where the HB becomes almost
vertical in the $V$ vs.\ $B-V$ plane. We will refer to the piece of the HB
between the blue edge of the RR Lyrae gap and this region as the blue HB
(BHB). Finally, we will refer to the part of the BT populated by stars
with $T_e\ge 23,000$~K as extreme horizontal branch (EHB). Observationally,
the beginning of the EHB region corresponds to the recently identified
``second jump'' (see below). 

\begin{figure}[h]
\vspace{1.8in}
\plotfiddle{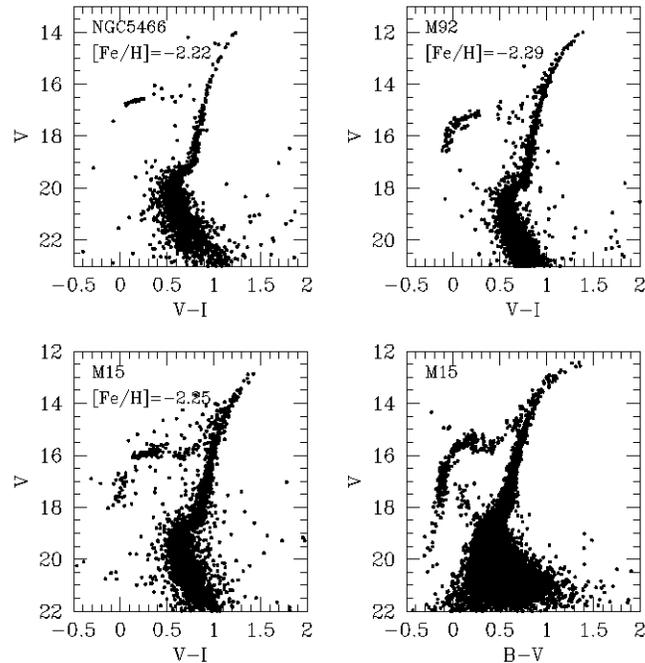}{6cm}{0}{60}{60}{-180}{-110}
\caption{ An example of the second parameter effect. The three clusters in 
the figure have the same metallicity within 0.1 dex (Harris 1996), but
show rather different CMDs, with the HB of M15 extending well below
the TO, and showing a gap. The $V$ vs.\ $V-I$ CMDs are from Rosenberg
et al.\ (2000), while the $V$ vs.\ $B-V$ CMD is from the HST snapshot
survey by Piotto et al.\ (2002).}
\end{figure}

%\vspace{-0.1in}

The blue tails probably represent the most extreme of the mixed bag of
anomalies that are sometime lumped into the term ``second-parameter
phenomena''. As an example, we can consider the HBs of NGC~5466, M92,
and M15 (Fig.~1).  Even if these three GCs have in practice the same
metallicity (within 0.1 dex, Harris 1996), they have completely
different HBs (Buonanno et al.\ 1985). NGC~5466 shows only a BHB. The M92 HB
extends beyond $T_e=10,000$~K, with a BT $\sim1$ magnitude long in the $V$
vs. $B-V$ or $V-I$ CMD, and no discontinuity in the stellar distribution.
M15 has a long BT, which extends well below the TO (as it can be
clearly seen in the more populated $V$ vs.\ $B-V$ CMD of the cluster core,
shown on the right side of Fig.\ 1), i.e., for more than 4 magnitudes,
reaching the EHB. Both in the $V$ vs.\ $V-I$ and in the $V$ vs.\ $B-V$ CMD,
there is a gap along the HB of M15, at $T_e\sim10,000$~K, i.e., exactly
where the HB of NGC~5466 seems to end.

In the last few years, a number of additional, unexpected, and mostly
not clearly understood observational facts about the horizontal branch
hot stars have come out, to further complicate the evolutionary scenario.

(1) It has been realized that HBs can extend well beyond
$T_e=30,000$~K (Sosin et al.\ 1997).  These stars are the GC
counterparts of the field blue subdwarfs (Newell 1973).  The hottest HB
stars lost most of their envelope during the red giant branch (RGB)
phase, and underwent a late helium core flash (D'Cruz et al.\ 1996)
while descending the white dwarf cooling sequence, with a deep mixing
(Brown et al.\ 2001) which might explain why these objects are less
luminous than predicted by canonical HB models. The high He content
and the large enhancement of carbon found in a sample of these objects
in $\omega$ Cen by Moehler et al.\ (2002) seems to support the
flash-mixing hypothesis of Brown et al.\ (2000).

(2) Thanks to the HST snapshot survey (Piotto et al.\ 2002), it has
been realized that BTs are present at all metallicities, including
the metal-rich clusters NGC~6388 and NGC~6441 (Rich et
al.\ 1997), for which canonical models would predict only a red clump.

(3) The BTs are present all over the cluster (Bedin et al.\ 2000), even
in its outskirts. This is a particularly intriguing result, as the presence
of BTs has been often attributed to an high density environment, in which 
close stellar encounters should favor tidal stripping, with a resulting 
shallower envelope, and therefore higher temperatures on the HB.

(4) Peterson et al.\ (2002) have found a large fraction of
binaries among the EHB stars.

(5) The BTs have a number of discontinuities, such as:
 
(i) The gaps, i.e., regions underpopulated with stars (Sosin et al.\ 1997,
see also Fig.\ 1). All the HBs with a BT in the snapshot sample (Piotto et
al.\ 2002) show a gap.  In some cases, these gaps might be simple
statistical fluctuations in the distribution of the stars along the HB
(Catelan et al.\ 1998). However, some gaps seem to appear at the same
temperature (Ferraro et al.\ 1998), and/or at the same mass (Piotto et
al.\ 1999) in all the BT clusters, pointing to some, common physical origin.

(ii) The first jump, at $T\sim 11,500$~K. Discovered for the
first time by Grundahl, VandenBerg, and Andersen (1998) in M13, this
discontinuity in the luminosity of the HBs in the Str\"omgren $u$,
$u-y$ CMD seems to be a feature present in all the BTs which extends
beyond $T_e=11,500$~K (Grundahl et al.\ 1999). It is also clearly seen
in the $U$, $U-V$ CMD (Fig.\ 3). 

(iii) The discontinuity in the surface gravities (Moehler et al.\ 2000,
and references therein).  In the interval $11,500 \le T_e \le 21,500$~K
spectroscopic observations suggest surface gravities smaller than
predicted by canonical models. 

\begin{figure}[h]
\vspace{0.1in}
\plotfiddle{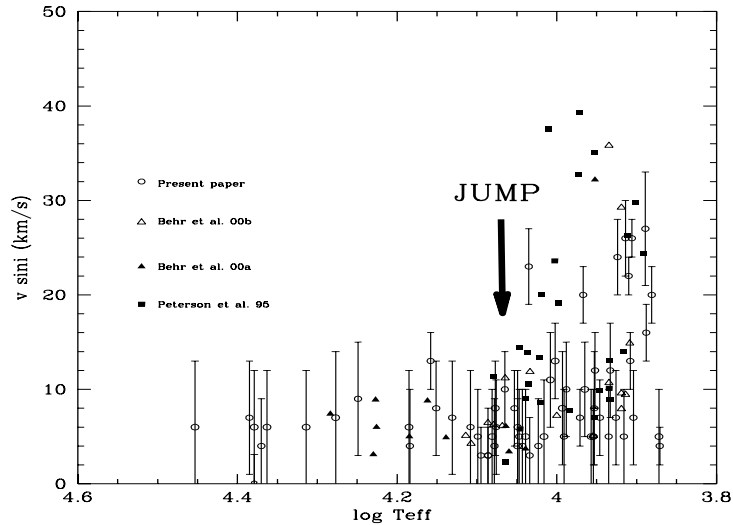}{6.0cm}{0}{50}{37}{-150}{-70}
\caption{ This figure shows all the projected rotation velocities of HB
stars measured so far in GCs. The {\it open circles} are the new data
from Recio-Blanco et al.\ (2002, 2003a, in prep., see also Recio-Blanco
et al.\ in this volume).  There is a clear discontinuity at the level of
the $u$-jump, at $T_e\sim 11,500$~K.}
\end{figure}

(iv) The discontinuity in the surface abundance ratios. Confirming
earlier results by Glaspey et al.\ (1989), Behr et al.\ (1999, 2000)
have shown that all the stars with $T_e>11,500$~K show an
overabundance of heavy elements (Fe, Ti, N, P, etc.)  with respect to
the cluster metallicity, and an under-abundance of helium. 

(v) The discontinuity in the stellar rotation rates (Behr et al.\ 2000,
Recio-Blanco et al.\ 2002). As shown in Fig.\ 2, the projected rotation
rates of BT stars show a sharp discontinuity at $T_e\sim11,500$~K
(Recio-Blanco et al.\ 2002), with the stars hotter than this
temperature being all slow rotators, while some star cooler than
$T_e=11,500$~K rotating faster than 40 km/s (see also Recio-Blanco et
al.\ in this volume for more details).  Recio-Blanco et al.\ (2002) have 
suggested that the jump in stellar rotation happens at the same
position of the photometric $u$-jump.

\begin{figure}[h]
\vspace{.1in}
\plotfiddle{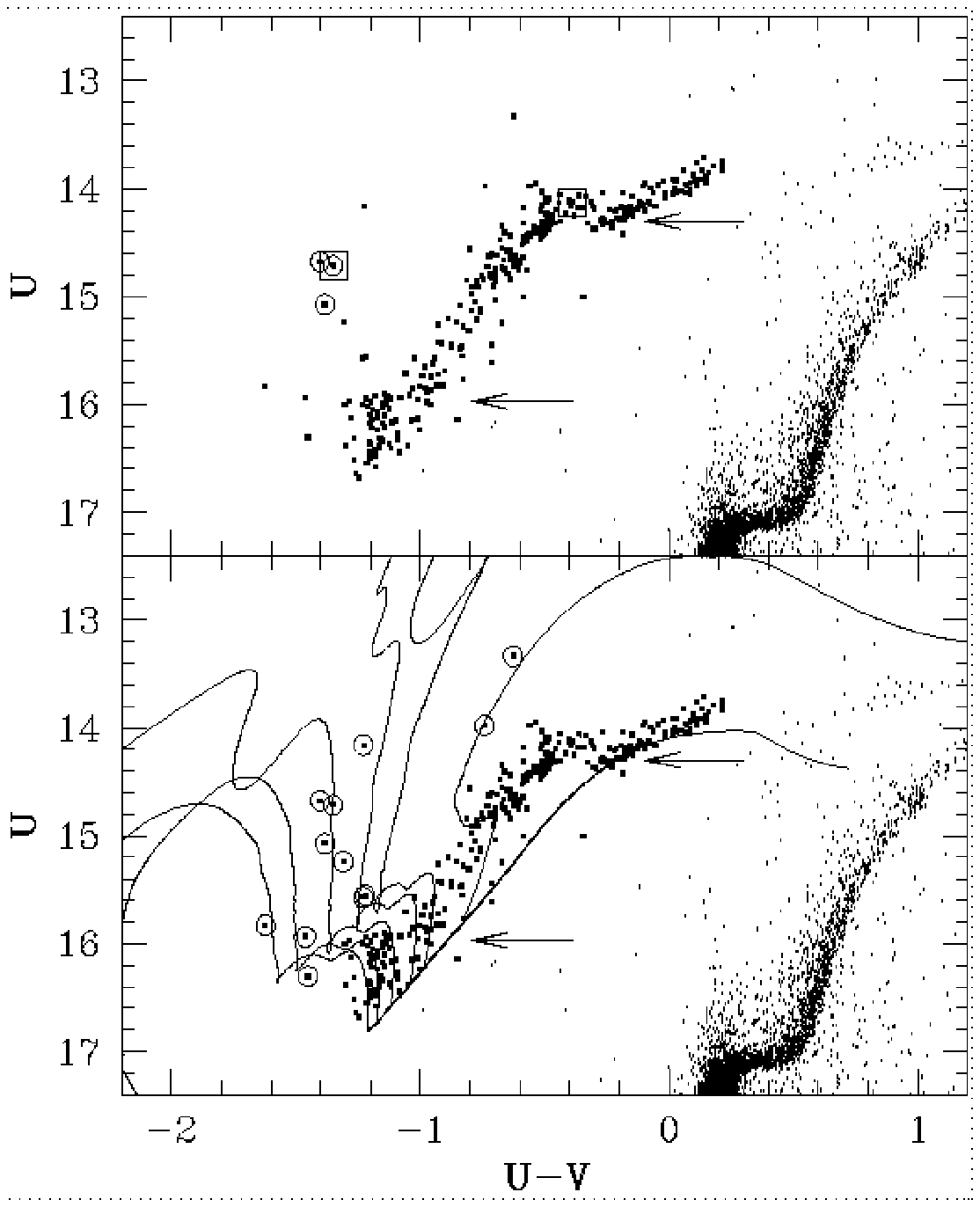}{8.5cm}{0}{50}{50}{-150}{-90}
\caption{{\it Upper panel:} The two jumps in the HB of NGC~6752, located at
$T_e\sim 11,500$~K (the $u$-jump by Grundahl et al.\ 1999) and at
$T_e\sim 23,500$~K (the ``second jump'' recently discovered by Momany
et al.\  (2002), respectively. The open circles and squares are post-HB
stars.  {\it Lower panel:} As in the {\it upper panel}, but with the
Z=0.0006 HB models superimposed. The {\it thick line} is the ZAHB,
while the {\it dotted lines} represent the off-ZAHB evolutionary
tracks. At the level of the ``second jump'' the evolutionary
tracks become almost vertical, while the evolution time becomes slower
and slower. Still, off-ZAHB evolution alone cannot explain this
``second jump''(cf.\ Momany et al.\ 2002 for more details).}
\end{figure}

(vi) The ``second jump'', at $T_e\sim23,500$~K. This feature (Fig.\ 3) has
been very recently discovered by Momany et al.\ (2002) in the $U$ vs.\ $U-B$
CMD of NGC~6752. The ``second jump'' seems to be an ubiquitous feature
(Momany et al.\ 2003, in prep.).  It is probably due to the
combination of post-zero-age HB evolution and diffusion effects.

It is clear that some of these discontinuities appear at approximately
the same temperature, and therefore might be the manifestation of the
same physical phenomenon. The abundance anomalies discovered by Behr
et al.\ (1999) are interpreted as observational evidence of the onset
of radiative levitation in stars hotter than $T_e=11,500$~K.
Following Greenstein, Truran, \& Cameron (1967), gravitational
settling of helium and radiative levitation of metals can occur in the
stable, non-convective atmospheres of the hot, high-gravity HB stars
(Michaud, Vauclair, \& Vauclair 1983). Grundahl et al.\ (1999) have
suggested that their $u$-jump can be the consequence of the metal
enhancement caused by radiative levitation. Higher abundances can also
(at least partially) account for the low-gravity problem affecting all
the stars with $T_e\ge11,500$ (Moehler et al.\ 2000).  The fact that
the change in the rotational velocity distribution of HB stars can be
associated with the $u$-jump (Fig.\ 2), makes the entire scenario
observationally consistent. The enhanced surface abundance of metals
can boost mass (and angular-momentum) loss via radiation pressure on
such elements, as suggested by Recio-Blanco et al.\ (2002).  Recent
models by Vink and Cassisi (2002) seem to confirm this scenario.

Despite these successes, we still do not understand the origin of the blue
tails, or of the gaps along them, and we do not know why fast rotators among
HB stars exist at all. However, the large photometric data bases that have
been collected in recent years (e.g., Rosenberg et al.\ 2000a,b, and Piotto
et al.\ 2002), and the even larger spectroscopic data bases that will
arrive in the near future thanks to the new multi-fiber facilities on the
10m-class telescopes will undoubtedly help to solve these long-standing
problems. An example of the kind of improvements we can expect is 
reported in the next Section.

\section{The Origin of BTs: New Observational Results}

One of the main targets of the HST snapshot project (Piotto et al.\
2002) was the investigation of the dependence of the GC HB morphology
on cluster parameters. The photometric quality, the large stellar population
sampled by the WFPC2 images in each cluster, and the large number (74) of
observed GCs provide a unique data base for this investigation. 

As shown by Djorgovski and Meylan (1994, see also Djorgovski in this volume), the
GCs represent a complex, multi-parameter family. The correct
statistical approach for the analysis of the relations among the many
parameters which define the GC family is a multivariate analysis. This
has been performed by Recio-Blanco (2003), and will be fully described
in a forthcoming paper by Recio-Blanco et al.\ (2003b, in prep.). Here
I want to anticipate a few new results, already evident in the simpler
univariate analysis, to show the potentialities of this database.

One of the problems we had to solve for this analysis was the definition
of a parameter able to describe the extension of the HB. There have been
many suggestions in the literature (cf.\ Piotto et al.\ 1999 and
references therein). Here we decided to use the effective
temperature of the hottest stars on the HB. This implies fitting the HB
with some theoretical models, but avoids all the problems that other
empirical parameters have (Catelan et al.\ 1998, Ferraro et al.\ 1998,
Piotto et al.\ 1999).

To fit a model to the HB we need a distance modulus and
a reddening.  In order to maintain internal consistency within the
snapshot database, we have remeasured all the distance moduli
and reddenings. The method will be fully described in Recio-Blanco et al.\ 
(2003c, in prep.). Here, suffice it to say that we adopted the same approach
as in Zoccali et al.\ (2000), though in this case we worked directly in the
HST F439W and F555W bands.

In the following, we will consider only the clusters with a BHB.  Once
the distance moduli and the reddenings had been determined, we fitted
ZAHB models specifically developed by Cassisi (2003, prov. comm.) to the observed
HBs.  This procedure allowed us to evaluate the highest effective
temperature (T$_{e}$) reached by the globular cluster HB and therefore
its temperature extent.  The errors in this temperature
determination are difficult to estimate, as they depend not only on the
errors in the distance modulus and reddening, but also
on the number of stars in the HB, and on the temperature range we have to
deal with. As a consequence, the largest errors occur for the
smallest, low-central-concentration clusters, and for the most extended
HBs, where the large bolometric correction in these photometric bands
precludes an accurate estimate of $T_{e}$.  However, although the
errors can be important, the general trend of the HB morphology with
cluster parameters is not dramatically affected.

Though we performed a full, multivariate analysis, we started with the
simplest univariate analyses.

The first correlation we identified is the HB morphology-metallicity
dependence. However, the large dispersion of the data clearly
indicates that there must be some other parameter which regulates the
HB extent.

\begin{figure}[h]
\vspace{.1in}
\plotfiddle{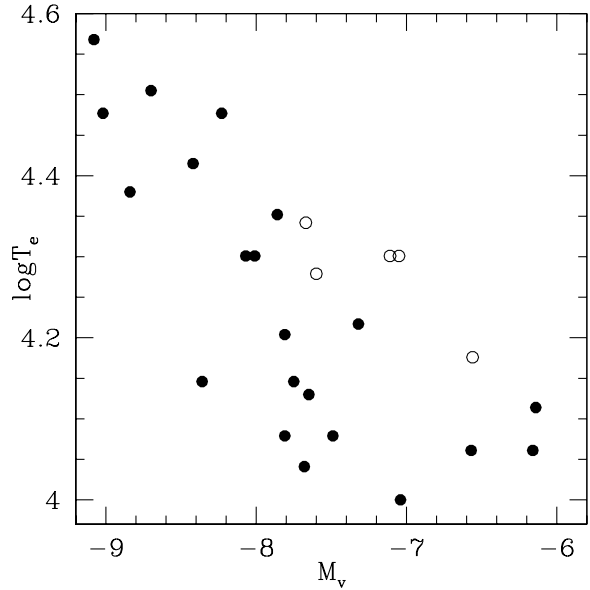}{7cm}{0}{40}{40}{-125}{-65}
\caption{The effective temperature of the hottest HB star is plotted against
the total absolute magnitude $M_V$ for a sample of
24 GCs with intermediate metallicity ($-1.90\le$~[Fe/H]~$\le-1.35$). The
{\it open circles} represent post-core-collapse clusters.}
\end{figure}

The most interesting results of this simple approach are summarized
in Figs. 4 and 5.  We isolated a subsample of 24 GCs with intermediate
metallicity in order to (i) remove as much as possible the metallicity
effect on the HB extent, and (ii) study the HB morphology 
in the metallicity interval where the second parameters are expected to
be most effective (Fusi Pecci et al.\ 1993). 

There is a significant correlation between the HB extension
and the total cluster luminosity ($M_V$) and the collisional parameter
($\Gamma_{\rm col}$).  The collisional parameter $\Gamma_{\rm col}$ is
defined as the number of collisions per unit time in the cluster core,
and it has been derived for a King-model cluster by King (2002):

\begin{center}

$\Gamma_{\rm col} = \log [5 \cdot 10^{-15} \sqrt{\sigma^3 \cdot r_c}]$

\end{center}

where $\sigma$   is  the  central   surface brightness  in  units   of
L$_\odot$/pc$^2$:
\begin{center}
  $\sigma = 10^{(-0.4 \cdot (\mu_V - 26.41))}$
\end{center}

\begin{figure}[h]
\vspace{.1in}
\plotfiddle{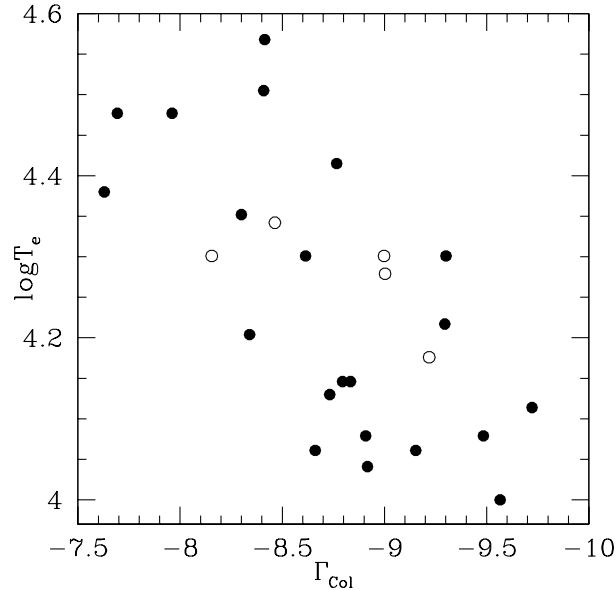}{7cm}{0}{40}{40}{-125}{-65}
\caption{The effective temperature of the hottest HB star is plotted against
the collisional parameter $\Gamma_{\rm coll}$ (King 2002) for the same sample of
GCs of Fig. 4. The
{\it open circles} represent post-core-collapse clusters.}
\end{figure}

Figure 4 shows that the most luminous clusters tend to
have the most extended HBs.  Noteworthy, at a given total luminosity,
the post-core-collapse (PCC) ({\it open circles}) clusters have the
most extended HBs: the PCC define the upper boundary of the relation
between $\log T_e$ and $M_V$. Luminosity is the most fundamental
observed quantity characterizing a stellar system, and for a set of
old stellar clusters it is a good relative measure of its baryonic
mass. Therefore, the observed trend suggests that more massive
clusters tend to have more extended HBs.

Figure 5 shows the correlation between the HB extension and the
collision rate. Clusters with higher probability of collisions among
stars in the core (the snapshot CMDs refer to the stars in the
cluster cores!), have, on average, more extended HBs. PCC clusters in
Fig.\ 5 seem to behave as normal King-model clusters, but this can be
due to the inadequacy of $\Gamma_{\rm col}$ in representing the
collision probability in a PCC cluster (indeed, $\Gamma_{\rm col}$ was
derived for a King-model cluster). 

Formally, the collisional parameter is independent of $M_V$, as it
has been calculated using the $\mu_V$ and r$_c$
parameters. However, we must expect some correlation between
$\Gamma_{\rm col}$ and $M_V$, because of the dependence of $M_V$ on
$\mu_V$ (Djorgovski and Meylan 1994).  In any case, multivariate
analysis (Recio-Blanco et al.\ 2003c) shows that the correlations in
Fig. 4 and 5 are both significant.  

It is tempting to interpret the correlation between the HB extent
and the collisional parameter as an indication that close stellar
encounters (or collisions) can cause some modification in the
evolution of the stellar population in GCs.  It is also worth to note
that the correlation between the HB extent and $\Gamma_{\rm col}$
is more significant than the correlation between the HB extent and
the central cluster density (Fusi Pecci et al.\ 1993), indicating that
$\Gamma_{\rm col}$ is a more appropriate parameter to describe the
collision effects in GC cores.

The dependence of the HB morphology on the total luminosity and
therefore the cluster total mass is also noteworthy. It cannot be a
simple consequence of the correlation of the HB extension with
$\Gamma_{\rm col}$.  One possible interpretation (Recio-Blanco 2003)
of the influence of M$_V$ and therefore, of cluster total mass, on the
HB morphology can be derived from some recent results by D'Antona et
al.\ (2002).  These authors analyze the consequences, on the HB
morphology, of helium variation due to self-pollution within GC
stars. Self-pollution had already been proposed as an explanation for
the chemical inhomogeneities (spread in the abundances of CNO, O--Na and
Mg anti-correlation) observed in GC members from the main sequence to the
RGB (e.g., Gratton et al.\ 2001).  The ejecta of massive asymptotic
giant branch stars, which would be the origin of the self-pollution,
would not only be CNO processed, but also helium enriched.  The models
of D'Antona et al.\ (2002) take into account this possible helium
enhancement.  They find that a spread in the helium content does not
affect the morphology of the main sequence, turn-off, and RGB in an
easily observable way, but that it can play a role in the formation of blue
tails. Indeed, structures with larger helium abundance are able to
populate much bluer HB regions. Self-pollution, and therefore helium
enrichment, are expected to be greater in more massive clusters, as they
would be able to retain the material from the ejecta better than less
massive ones.  Therefore, we should expect more extended HBs in more
massive clusters, as we observe. It would be extremely interesting to
investigate whether the clusters with the most extended HB also show a Na
and Mg anti-correlation, which would confirm this scenario (D'Antona et
al.\ 2002).

In any case, helium variation due to self-pollution is not enough to
explain the observed HB morphology.  For example, higher mass loss is
needed in metal-rich clusters than in metal-poor ones, to produce
bluer HB stars. Close stellar encounters, in
clusters with higher collision rates, contribute to explaining the complex
observational scenario, as suggested by Fig.\ 5.

Another noteworthy example of the effect of the environment on the
evolution of the stellar population of GCs is discussed by De Angeli
and Piotto, in this volume. Using the same snapshot data base that has
been used here, we find a dependence of the relative frequency of
blue stragglers on both the total cluster luminosity and the collision rate.
In particular, we find that clusters with higher collision rates show
a smaller relative number of blue stragglers. We also find that the most
luminous clusters have the smallest population of BSS. 

\section{Conclusions}

There was a large debate in 1992, at the first Ivanfest, on the
interplay between dynamical evolution and stellar evolution (Djorgovski
and Piotto 1993, and other articles in the same volume).  At that time
the observational evidence was scarce and mostly contradictory.  After
10 years, it seems we have collected enough observational information,
mostly from photometric data, to confirm this effect.  We hope that in
ten years from now we can meet again, feasting Ivan's 85th birthday,
after having digested the huge amount of photometric data that we have
collected with the recently developed wide field imagers, and the even
larger data set that we expect to collect with the forthcoming
multi-fiber spectroscopic facilities. It is easy to foresee that by that
time, thanks to these observational inputs, and with the help of
parallel improvements of the models, we will have a much deeper
knowledge of the evolution processes at the basis of the formation of
the complex and still not completely understood horizontal branches.

\begin{acknowledgments} 
I thank my collaborators in Padova, Luigi Bedin,\break Francesca De Angeli,
Yazan Momany, and Alejandra Recio-Blanco, for their hard work and their
enthusiasm in carrying out the many projects we started, and that are in
part described in this paper. I also thank Santino Cassisi and
Giuseppe Bono for their \dots ``theoretical support''. I thank Ivan King
for reading and correcting this manuscript. This project has
been partially supported by the Agenzia Spaziale Italiana and by the
Ministero dell'Istruzione, Universit\`a e Ricerca under the program
``Cofin2001''. 
\end{acknowledgments}

\end{document}